\documentclass[12pt]{article}

\usepackage{graphicx}
\usepackage{amsfonts}
\usepackage[english]{babel}
\usepackage[utf8]{inputenc}
\usepackage{amsthm}
\usepackage{amssymb, amsmath, pxfonts, enumitem, mathtools}
\usepackage{mathrsfs}
\usepackage{graphicx}
\usepackage{caption}
\usepackage{subcaption}

\input{epsf}

\def\br{\begin{eqnarray}}
\def\er{\end{eqnarray}}
\def\be{\begin{equation}}
\def\ee{\end{equation}}
\def\({\left(}
\def\){\right)}

\def\rlx{\relax\leavevmode}

\newcommand{\norm}[1]{\Vert {#1} \Vert}

\def\IZ{\rlx\hbox{\sf Z\kern-.4em Z}}
\def\IR{\rlx\hbox{\rm I\kern-.18em R}}
\def\IC{\rlx\hbox{\,$\inbar\kern-.3em{\rm C}$}}
\def\one{\hbox{{1}\kern-.25em\hbox{l}}}

\begin{document}

\begin{titlepage}
\vspace*{-1cm}

\vskip 2cm

\vspace{.2in}
\begin{center}
{\large\bf BPS states for scalar field theories based on $\mathfrak{g}_2$ and $\mathfrak{su}(4)$ algebras}
\end{center}

\vspace{.5cm}

\begin{center}
G. Luchini~$^{\dagger}$ and T. Tassis$^{\dagger}$

\vspace{.3 in}
\small

\par \vskip .2in \noindent 
$^{\dagger}$ Departamento de F\'isica\\
 Universidade Federal do Esp\'irito Santo (UFES),\\
 CEP 29075-900, Vit\'oria-ES, Brazil

\vspace{2cm}

\normalsize
\end{center}

%\vspace{.5in}

\begin{abstract}

\noindent We discuss two models in $1+1$ dimensional space-time for real scalar field multiplets on the root space of $\mathfrak{g}_2$ and $\mathfrak{su}(4)$ Lie algebras. The construction of these models is presented and the corresponding BPS solutions are found.

\end{abstract} 

\end{titlepage}

\section{Introduction}
\label{sec:intro}
\setcounter{equation}{0}

The concept of soliton has appeared in the study of the famous KdV equation\cite{doi:10.1080/14786449508620739}\cite{PhysRevLett.15.240} as a wave solution which is localised in space and time, that keeps its shape during propagation and when undergoing scattering processes. These features led to the interpretation of solitons as particle-like objects even as classical fields. 

Soliton solutions are commonly found in integrable field theories\cite{das1989integrable}\cite{Lax1968IntegralsON} and their stability is associated to the existence of infinitely many conservation laws\cite{Ferreira:2007pr}. How to extend the concept of integrability for higher dimensional space-times so that one can use in this context all the very robust and powerful algebraic techniques\cite{Zakharov:1979zz} in the construction of soliton solutions as it is done in low dimensions is yet to be understood\cite{Alvarez:2009dt}. Nonetheless, the nonperturbative treatment of field theories in higher dimensions leads to solutions which share many of the features of solitons\cite{Manton_2008} having their stability now defined by the topological character of the field; these are the topological solitons\cite{manton2004topological}.

As a fundamental ingredient, a topological soliton has a topological degree that is defined by the pullback of the volume form of the target space into the space manifold. This quantity, by construction, is completely defined by the topological data, which is determined by the behavior of the field at the border of space. This is a crucial criterium for the finiteness of the energy, which is the basic feature of the solitonic solution. In some theories one may find through the Bogomolny bound\cite{Bogomolny:1975de} a crucial relation between the topological degree and the static energy; not only that, but also one finds a simpler static equation which determines the configuration with least possible energy, known as the BPS state. 

Here we discuss two models which have BPS states derived from a class of models recently proposed\cite{Ferreira:2018ntx} by Ferreira, Klimas and Zackrzewsky, which are referred here as FKZ models. They are relativistic theories for a real scalar field multiplets in $1+1$ dimensional space-time where the target or field space is defined to be that of the roots of the Lie algebras, which here are taken to be $\mathfrak{g}_2$ and $\mathfrak{su}(4)$ respectively. In section \ref{sec:review} we briefly review the formulation of the generalized BPS equations which leads to the construction of the FKZ models, described in the same section. Then, in section \ref{sec:g2} we discuss the construction of the model for the algebra $\mathfrak{g}_2$ and present some of its BPS states obtained numerically. In doing so, we analyse some aspects of the potential and some important features which are presented in this type of model, such as the exitence of an infinity of BPS states and the possibility of finding interesting submodels. In section \ref{sec:su4} we discuss the construction of the FKZ model based on the $\mathfrak{su}(4)$ algebra. This is a model for a scalar triplet; while in \cite{Ferreira:2018ntx} algebras of rank 2 only were considered, here we show a FKZ model for a rank 3 algebra. In this case, we have found the possibility of a continuous vacuum manifold, which does not occured for the rank 2 algebras based FKZ models.

\section{A review of the generalized BPS equation and the FKZ models}
\label{sec:review}
\setcounter{equation}{0}

\subsection{The generalization of the BPS equation for scalar field multiplets}
\label{subsec:genbps}
\setcounter{equation}{0}

The so called BPS equations\cite{Bogomolny:1975de}\cite{Prasad:1975kr} are of great importance in the construction of static solitonic solutions in many different nonlinear topological field theories, in the context of vortices\cite{deVega:1976xbp}, monopoles\cite{Sutcliffe:1997ec} and domain walls. The remarkable feature of these equations lies on the fact that they are of first order with respect to spatial derivatives and they imply the usual second order static equations of the theory\footnote{Which is assumed here to be described by a lagrangian that has terms that, at most, depend on the first derivatives of the field}. Not only that but configurations which satisfy the BPS equations, the so called BPS states, are those for which the energy has the least possible value, being proportional to the least topological degree of the field\cite{manton2004topological}. The establishment of the BPS equation follows the procedure known as the ``Bogomolny trick'', consisting essentially of a redefinition of the static energy such that a relation between the first derivative of the field and the field itself appears explicitly.

Some years ago some aspects of the BPS equation were discussed in \cite{Adam:2013hza}. The authors argued that the BPS equation is the condition that a field configuration must satisfy in order to be both topological, \textit{i.e.} to have a topological charge associated to it, and to leave a specific functional of the field and its first derivatives (which can coincide with the static energy functional of that theory) stationary. This observation leads not only to a satisfactory explanation on why the ``Bogomolny trick'' works but also it gives a fundamental recipe to the construction of new BPS models\cite{Bazeia:1995en}\cite{Bazeia:2018srv} \cite{Bazeia:2017cqv}.  

In $1+1$ dimensional space, for a single real scalar field $\phi$, the argument goes as follows. One starts by considering the quantities $A(\phi,\phi')$ and $\widetilde{A}(\phi,\phi')$ which are functions of the field $\phi$ and its spatial derivative $\phi'\equiv \frac{d\phi}{dx}$. Then, it is claimed that the solution of the static equation of motion is a configuration which leaves the functional
\begin{equation}
\label{eq:staticengy}
E = \frac{1}{2}\int_{-\infty}^{+\infty}dx\;\left(A^2 + \widetilde{A}^2\right)
\end{equation}
stationary at first order under variations of the field $\phi \rightarrow \phi + \delta \phi$ with $\delta \phi = 0$ at $x=\pm \infty$.
The stationarity of $E$ imposes a condition on the functions $A$ and $\widetilde{A}$ which is
\begin{equation}
\label{eq:varengy}
A\frac{\delta A}{\delta \phi}-\frac{d}{dx}\left(A\frac{\delta A}{\delta \phi'}\right) + (A \leftrightarrow \widetilde{A})=0.
\end{equation}

Next, one also assumes that there is a quantity associated to the field configuration defined as
\begin{equation}
\label{eq:topcharge}
Q = \int_{-\infty}^{+\infty}dx\;A\widetilde{A}
\end{equation}
which remains unchanged under this same kind of variation. The invariance of $Q$ under these tranformations of the field implies the following conditions over $A$ and $\widetilde{A}$:
\begin{equation}
\label{eq:vartopcharge}
\widetilde{A}\frac{\delta A}{\delta \phi}-\frac{d}{dx}\left(\widetilde{A}\frac{\delta A}{\delta \phi'}\right)+(A \leftrightarrow \widetilde{A})=0.
\end{equation}

It is then not difficult to see that the relation
\begin{equation}
\label{bps}
A = \pm \widetilde{A}
\end{equation}
defines a compatibility condition between equations (\ref{eq:varengy}) and (\ref{eq:vartopcharge}). This is a relation between $\phi'$ and $\phi$, recognised as the BPS equation associated to the theory described by the lagrangian
\begin{equation}
\label{eq:1scalarlag}
\mathcal{L} = \frac{1}{2}\partial_\mu \phi \partial^\mu \phi - \mathcal{U}(\phi),
\end{equation}
if one defines
\begin{equation}
A = \frac{d\phi}{dx} \qquad\textrm{and}\qquad \widetilde{A} \equiv \frac{dW}{d\phi} = \sqrt{2\mathcal{U}},
\end{equation}
thus $E$ is equivalent to the static energy
\begin{equation}
E = \int_{-\infty}^{+\infty}dx\; \left(\frac{1}{2}\left(\frac{d\phi}{dx}\right)^2+\mathcal{U}\right)
\end{equation}
and 
\begin{equation}
Q = \int_{-\infty}^{+\infty}\frac{d\phi}{dx}\frac{dW}{d\phi} = W(\phi(+\infty)) - W(\phi(-\infty))
\end{equation} 
has clearly a topological character, depending only on the value of the function $W$ referred to as the prepotential, evaluated on the values of the field at spatial infinity, which defines the topological data of the model.

Moreover, from (\ref{eq:staticengy}), one can easily see how the usual ``Bogomolny trick'' is performed, giving rise to both, the BPS equation and the minimum energy value, equal to the above defined topological charge $Q$:
\begin{equation}
\label{eq:bogotrick}
E = \frac{1}{2}\int_{-\infty}^{+\infty}dx\;\left(A\mp \widetilde{A}\right)^2 \pm Q \geq Q. 
\end{equation}

Next, one can extend this construction to a multiplet of scalar fields with components $\phi^a \in \mathbb{R}$, $a=1,2,\dots,n$ in a field space with metric $\eta_{ab}$, described by the lagrangian
\begin{equation}
\label{eq:lag_field}
\mathcal{L}= \frac{1}{2}\eta_{ab}\partial_\mu \phi_a\partial^\mu\phi_b - \mathcal{U}.
\end{equation}

For constant metric components the dynamical equation of this model reads
\begin{equation}
\label{eq:eom}
\eta_{ab}\partial_\mu \partial^\mu \phi_b +\frac{\partial \mathcal{U}}{\partial \phi_a}=0.
\end{equation}

Through an analogous procedure as done before for a single field, one defines\cite{Adam:2013hza} the functionals $Q$ and $E$ as
\begin{equation}
Q = \int_{-\infty}^{+\infty}dx\;A_a \widetilde{A}_a \qquad \textrm{and}\qquad E = \frac{1}{2}\int_{-\infty}^{+\infty}dx\;\left(A_a^2 + \widetilde{A}_a^2\right)
\end{equation}

where now $A$ and $\widetilde{A}$ are vectors in field space defined as
\begin{equation}
A_a = k_{ab}\frac{d\phi^b}{dx} \qquad \qquad \widetilde{A}_a = \frac{\partial W}{\partial \phi_b}k_{ba}^{-1}
\end{equation}
with $k$ being related to the field space metric $\eta_{ab}$ by $\eta = k^Tk$ and the prepotential $W$ given in terms of the potential energy density $\mathcal{U}$ by 
\begin{equation}
\label{eq:prepotential}
\mathcal{U}=\frac{1}{2}\eta_{ab}^{-1}\frac{\partial W}{\partial \phi_a}\frac{\partial W}{\partial \phi_b}.
\end{equation}

The BPS equation for this theory is defined as the compatibility condition for the stationarity of the functional $E$ and the invariance of $Q$ under local variations of the field, and it reads
\begin{equation}
\label{eq:bpsmultiplet}
\frac{d\phi_a}{dx}=\eta_{ab}^{-1}\frac{\partial W}{\partial \phi_b}.
\end{equation}

The BPS solutions thus have topological charge defined by
\begin{equation}
\label{eq:topbps}
Q = \int_{-\infty}^{+\infty} d\phi \cdot \nabla_\phi W = W(\phi(+\infty))-W(\phi(-\infty))
\end{equation}

where $\nabla_\phi W$ stands for the gradient of the prepotential in field space: $(\nabla_\phi W)_a = \frac{\partial W}{\partial\phi_a}$.

The BPS states are those which interpolate between vacua of the theory with least possible energy, the vacua being then defined by the minima of the potential. In terms of the prepotential, the vacua are characterized by its points of extrema\cite{Ferreira:2018ntx}, either maxima or minima:
\begin{equation}
\label{eq:vaccond}
\frac{\partial W}{\partial\phi_a}=0.
\end{equation}

The static field configuration $\phi(x)$ can be seen as a path in the field space parameterized by the coordinate $x$. Then, $v_a\equiv \frac{d\phi_a}{dx}$ is the velocity vector of this path, tangent to it everywhere. The BPS equation (\ref{eq:bpsmultiplet}) can thus be written as
\begin{equation}
v = \pm \nabla_\eta W
\end{equation}
where $(\nabla_\eta W)_a\equiv \eta_{ab}^{-1}\frac{\partial W}{\partial \phi_b}$, following \cite{Ferreira:2018ntx} is called the $\eta$-gradient of $W$. Thus, the BPS equation as written above tell us that the vector that is tangent to the BPS state is equal, at each point, to the $\eta$-gradient of the prepotential $W$, that is, each solution to the BPS equations is given by a path following the $\eta$-gradient lines of $W$. 

The paths of BPS states never intersect each other since this would mean that the $\eta$-gradient of $W$ is multivalued. The $\eta$-gradient lines can at most meet tangentially or converge to points where $\nabla_\eta W  = 0$, \textit{i.e.} the vacua of the potential energy density $\mathcal{U}$ are sources or sinks of $\eta$-gradient lines. Since the finite energy BPS states start and finish at vacua points, this means that the paths they describe in field space connect a source to a sink of $\eta$-gradient lines. This fact implies that the prepotential $W$ varies monotonically across the path of a given configuration.

%%%%%%%%%%%%%%%%%%%%%%%%%%%%%%%%%%%%%%%%%%%%%%%%%%%%%%%%%

\subsection{The FKZ models}

The ideas very briefly reviewed above were first discussed in \cite{Adam:2013hza} and applied in the study of different models since then. The method that leads to the generalization of the BPS equation is quite general and not only gives an explanation to the existense of some known BPS equations but also can be used in a straightforward manner in formulating new field theoretical BPS models. In this section we shall review a class of such models, refered to here as the FKZ models, introduced recently by Ferreira, Klimas and Zakcrzewski in \cite{Ferreira:2018ntx}.

In this construction, the scalar multiplet $\varphi = (\phi_1, \ldots, \phi_r)$ is defined in the space spanned by the simple roots $\alpha_a$, $a = 1, \ldots, r$ of a given Lie algebra $\mathfrak{g}$ of rank $r$:
\begin{equation}\label{eq:def phi}
\varphi \equiv \sum_{a = 1}^{r} \phi_a \frac{2 \; \alpha_a}{\norm{\alpha_a}^2}.
\end{equation}

The field space metric $\eta_{ab}$ is then defined by the Cartan matrix $K_{ab}= 2\frac{\alpha_a\cdot \alpha_b}{\norm{\alpha_b}^2}$ of $\mathfrak{g}$:
\begin{equation}
\eta_{ab}= \frac{2K_{ab}}{\norm{\alpha_a}^2}.
\end{equation}

The construction of these models is based on the prepotential $W$, from where the potential is defined using (\ref{eq:prepotential}). In the case of the FKZ models, $W$ is a scalar in the space of simple roots. One chooses a representation $\mathcal{R}$ of $\mathfrak{g}$ with weights $\mu_k$ and define
\begin{equation}
W \equiv \sum_{\mu_k \in \mathcal{R}} C_{\mu_k} \; e^{i \mu_k \cdot \varphi}.
\end{equation}

For a real prepotential, that is, $W = W^\ast$, one considers representations satisfying $ \mu_k \in \mathcal{R} \Leftrightarrow - \mu_k \in \mathcal{R}$ with which it is possible to write
\begin{equation}
W = \sum_{\mu_k \in \mathcal{R}^{(+)}} \left( C_{\mu_k} \; e^{i \mu_k \cdot \varphi} + C_{-\mu_k} \; e^{i (-\mu_k) \cdot \varphi} \right)
\end{equation}
where $\mathcal{R}^{(+)}$ stands for the fact that only one weight $\mu_k$ out of each pair $(\mu_k, -\mu_k) \in \mathcal{R}$ is considered\footnote{Notice that if the representation has weights with value zero that would only add a constant in the prepotential and since we are only interested in derivatives -- or differences -- of the prepotential, we can always ignore additive constants.}. And so the reality condition implies that the coefficients have to satisfy $C_{\mu_k} = C^*_{-\mu_k}$. Defining $C_{\mu_k} \equiv \frac{1}{2} \left( \gamma_{\mu_k} - i \delta_{\mu_k} \right)$, such that $\gamma, \delta \in \mathbb{R}$ and $\gamma_{\mu_k} = \gamma_{-\mu_k} \textrm{ , } \; \delta_{\mu_k} = - \delta_{-\mu_k}$
then one finally has
\begin{equation}
W = \sum_{\mu_k \in \mathcal{R}^{(+)}} \left[ \gamma_{\mu_k} \cos(\mu_k \cdot \varphi) + \delta_{\mu_k} \sin(\mu_k \cdot \varphi) \right].
\end{equation}

This expression gives the general form of the prepotential for the FKZ models, as presented in \cite{Ferreira:2018ntx}. These theories can become fairly complicated very fast. One allows itself a bit of simplification and considers only models for which $\delta_{\mu_k} = 0$, \emph{i.e.}, only prepotentials of the form
\begin{equation} \label{eq:superpot}
W = \sum_{\mu_k \in \mathcal{R^{(+)}}} \gamma_{\mu_k} \cos(\mu_k \cdot \varphi).
\end{equation}
Even with this restriction the models that emerge are very rich. The potential for these theories will have the general form given in (\ref{eq:prepotential}) with
\begin{equation}
\frac{\partial W}{\partial \phi_a} = - 2 \frac{\mu_k \cdot \alpha_a}{\norm{\alpha_a}^2} \sum_{\mu_k \in \mathcal{R^{(+)}}} \gamma_{\mu_k} \sin \left(2 \sum_b \phi_b \frac{\mu_k \cdot \alpha_b}{\norm{\alpha_b}^2} \right).
\end{equation}
Further, the vacua will satisfy equation (\ref{eq:vaccond}), that in this case reads
\begin{equation} \label{eq: fkz vac}
\frac{\mu_k \cdot \alpha_a}{\norm{\alpha_a}^2} \sum_{\mu_k \in \mathcal{R^{(+)}}} \gamma_{\mu_k} \sin \left(2 \sum_b \phi_b \frac{\mu_k \cdot \alpha_b}{\norm{\alpha_b}^2} \right) = 0.
\end{equation}
The vacuum manifold structure will be, in general, very complex and depend heavily on the values of the constants $\gamma_k$. In particular, the points
\begin{equation}
\phi_a = n_a \pi,
\end{equation}
for $n_a \in \mathbb{Z}$, will always be in the vacuum set since from Lie algebra theory one has that the weights $\mu_k$ always satisfy $\frac{2 \mu_k \cdot \alpha_a}{\norm{\alpha_a}^2} = m_{ka}$, where $m_{ka} \in \mathbb{Z}$, so that
\begin{equation}
\sin \left( 2 \pi \sum_a m_{ka} n_a \right) = 0,
\end{equation}
as the sum of integers is an integer. Other types of vacua, that rely on further properties of Lie algebra theory are discussed in \cite{Ferreira:2018ntx}.

In \cite{Ferreira:2018ntx} the authors present models based on the algebras $\mathfrak{su}(2)$, $\mathfrak{su}(3)$ and $\mathfrak{so}(5)$. Here we shall discuss the cases for the algebras $\mathfrak{g}_2$ and $\mathfrak{su}(4)$, theories with two and three fields, respectively, and some of their static solutions.

%%%%%%%%%%%%%%%%%%%%%%%%%%%%%%%%%%%%%%%%%%%%%%%%%%%%%%%%%%%%%%%%%%%%%%%%%%%%%%%

\section{The FKZ model for the algebra $\mathfrak{g}_2$}
\label{sec:g2}
\setcounter{equation}{0}

\subsection{The construction of the model}

The algebra $\mathfrak{g}_2$ is of rank $r = 2$ and following the FKZ prescription, it is suitable for the description of scalar doublet with components $\phi_1$ and $\phi_2$:
\begin{align}
\varphi = \phi_1 \frac{2 \alpha_1}{\norm{\alpha_1}^2} + \phi_2 \frac{2 \alpha_2}{\norm{\alpha_2}^2}.
\end{align}
The Cartan matrix of this algebra is given by 
\begin{align}
K = \left( \begin{array}{cc}
2	& -1 \\
-3 	& 2
\end{array} \right).
\end{align}
Using the fact that the norms of the simple roots satisfy $\frac{\norm{\alpha_a}^2}{\norm{\alpha_b}^2} = \frac{K_{ab}}{K_{ba}}$, we choose the normalization $\norm{\alpha_1}^2 = 1$ and $\norm{\alpha_2}^2 = 3$, so that the matrix of the field space reads
\begin{align}
\eta = \left( \begin{array}{cc}
4 & -2 \\
-2 & 4/3
\end{array} \right).
\end{align}

The next step in the construction of the model is to choose a representation for this algebra. A good starting point is to consider the fundamental representations, \emph{i.e.}, representations for which the highest weight is a fundamental one. For an algebra of rank $r$, the fundamental weights are given by
\begin{align}
\lambda_a = \sum_{b = 1}^r K_{ab}^{-1} \alpha_b
\end{align}
where $K^{-1}$ is the inverse of the Cartan matrix. In the case of $\mathfrak{g}_2$, there are two fundamental weights
\begin{align}
\lambda_1 = 2 \alpha_1 + \alpha_2 \quad \text{ and } \quad \lambda_2 = 3 \alpha_1 + 2 \alpha_2.
\end{align}
The first fundamental representation, \emph{i.e.}, the representation with highest weight $\lambda_1$, has the following weights
\begin{align}
\mu_1 &= \lambda_1 = 2 \alpha_1 + \alpha_2 \nonumber \\
\mu_2 &= \lambda_1 - \alpha_1 = \alpha_1 + \alpha_2 \nonumber \\
\mu_3 &= \lambda_1 - \alpha_1 - \alpha_2 = \alpha_1 \nonumber \\
\mu_4 &= \lambda_1 - 2 \alpha_1 - \alpha_2 = 0 \\
\mu_5 &= \lambda_1 - 3 \alpha_1 - \alpha_2 = - \alpha_1 = - \mu_3 \nonumber \\
\mu_6 &= \lambda_1 - 3 \alpha_1 - 2 \alpha_2 = - \alpha_1 - \alpha_2 = -\mu_2 \nonumber \\
\mu_7 &= \lambda_1 - 4 \alpha_1 - 2 \alpha_2 = - 2 \alpha_1 - \alpha_2 = - \mu_1 \nonumber
\end{align}
and they already satisfy the requirement for the reality of the prepotential $W$. 

In order to calculate the internal products $\mu_k \cdot \varphi$, which are passed as arguments for the cosines in the prepotential, one uses the orthogonality relation between simple roots and the fundamental weights, $\frac{2 \lambda_a \cdot \alpha_b}{\norm{\alpha_b}^2} = \delta_{ab}$, obtaining
\begin{align}
\mu_1 \cdot \varphi &= \sum_a \phi_a \frac{2 \lambda_1 \cdot \alpha_a}{\norm{\alpha_a}^2} = \sum_a \phi_a \delta_{1a} = \phi_1 \nonumber \\
\mu_2 \cdot \varphi &= \sum_a \phi_a \frac{2 (\lambda_1 - \alpha_1) \cdot \alpha_a}{\norm{\alpha_a}^2} = \phi_1 - K_{11} \phi_1 - K_{12} \phi_2 = - \phi_1 + \phi_2 \\
\mu_3 \cdot \varphi &= \sum_a \phi_a \frac{2 (\lambda_1 - \alpha_1 - \alpha_2) \cdot \alpha_a}{\norm{\alpha_a}^2} = \phi_1 - K_{11} \phi_1 - K_{12} \phi_2 - K_{21} \phi_1 - K_{12} \phi_2 = 2 \phi_1 - \phi_2 \nonumber
\end{align}
and finally the prepotential reads
\begin{equation}
\label{eq:g2-prepot}
W = \gamma_1 \cos \phi_1 + \gamma_2 \cos(\phi_1 - \phi_2) + \gamma_3 \cos(2 \phi_1 - \phi_2).
\end{equation}

The components of the gradient of $W$ in field space are
\begin{align}
\frac{\partial W}{\partial \phi_1} &= - \gamma_1 \sin \phi_1 - \gamma_2 \sin(\phi_1 - \phi_2) - 2 \gamma_3 \sin(2 \phi_1 - \phi_2) \nonumber \\
\frac{\partial W}{\partial \phi_2} &= \gamma_2 \sin(\phi_1 - \phi_2) + \gamma_3 \sin(2 \phi_1 - \phi_2).
\end{align}
and the potential becomes
\begin{eqnarray} \label{eq: g2 pot}
U(\phi) &=& \frac{1}{2} \left[ \left( \frac{\partial W}{\partial \phi_1} \right)^2 + 3 \frac{\partial W}{\partial \phi_1} \frac{\partial W}{\partial \phi_2} + 3 \left( \frac{\partial W}{\partial \phi_2} \right)^2 \right]\nonumber\\
&=&\frac{1}{2} \left[\gamma _1^2 \sin ^2\left(\phi _1\right)+\gamma _2^2 \sin ^2\left(\phi _1-\phi _2\right)+\gamma _3^2 \sin ^2\left(2 \phi _1-\phi _2\right)-\gamma _2 \gamma _1 \sin \left(\phi _1\right) \sin \left(\phi _1-\phi _2\right)+\nonumber\right.\\
&+&\left.\gamma _3 \sin \left(2 \phi _1-\phi _2\right) \left(\gamma _1 \sin \left(\phi _1\right)+\gamma _2 \sin \left(\phi _1-\phi _2\right)\right)\right].
\end{eqnarray}

\subsection{BPS solutions}

The first step in determining the BPS solutions of this model is the definition of the vacua of the potential, which are given by the critical points of the prepotential, that is, the points $\varphi_0 = (\phi_1, \phi_2)$ which satisfy
\begin{align} \label{eq: g2 vac cond}
\gamma_1 \sin \phi_1 + \gamma_2 \sin(\phi_1 - \phi_2) + 2 \gamma_3 \sin(2 \phi_1 - \phi_2) &= 0 \nonumber \\
\gamma_2 \sin(\phi_1 - \phi_2) + \gamma_3 \sin(2 \phi_1 - \phi_2) &= 0.
\end{align}
This set will depend on the choice of the parameters $\gamma_i$. For the particular choice $\gamma_i = 1$, $i=1,2,3$, the vacua are
\begin{align}
\varphi_0=\left\{
\begin{array}{c}
(n_1\pi, n_2\pi)\\
(\frac{2\pi}{3}+2\pi n_1, 2\pi n_2)\\
(\frac{4\pi}{3}+2\pi n_1, 2\pi n_2)
\end{array}
\right.
\end{align}
where $n_1, n_2 \in \mathbb{Z}$. In figure \ref{fig: g2-pot} we present the plots of the above potential and prepotential for this choice of $\gamma_i$.
\begin{figure}[h!]
	\centering
	\begin{subfigure}{0.49\textwidth} % width of left subfigure
		\includegraphics[width=\textwidth]{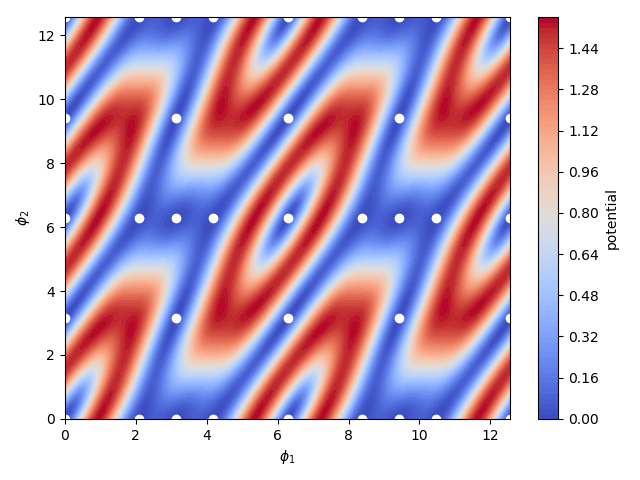}
		\caption{The plot of the potential given in (\ref{eq: g2 pot}).} % subcaption
		\label{fig: g2-pot}
	\end{subfigure}
	\vspace{1em} % here you can insert horizontal or vertical space
	\begin{subfigure}{0.49\textwidth} % width of right subfigure
		\includegraphics[width=\textwidth]{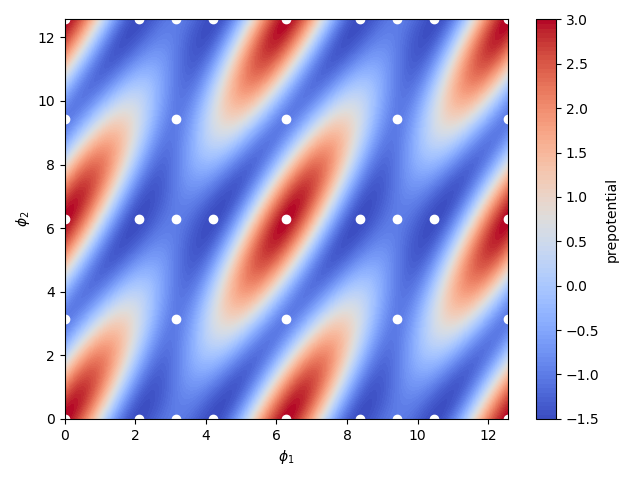}
		\caption{The plot of the prepotential defined in (\ref{eq:g2-prepot}).} % subcaption
		\label{fig: g2-prepot}
	\end{subfigure}
	\caption{The plots of the potential and of the prepotential for the $\mathfrak{g}2$ algebra FKZ model. The corresponding discrete vacua are defined at the points on the plane $\phi_1-\phi_2$ with white dots. The vacua are seen to be localised at the points of extremum of the prepotential which always correspond to points of minima of the potential.} % caption for whole figure
\label{fig: g2-pot}
\end{figure}

The BPS equations for the model then read
\begin{eqnarray}
\label{eq: bps-g2}
%\frac{d \phi_1}{d x} &= \pm \left( \eta_{11}^{-1} \frac{\partial W}{\partial \phi_1} + \eta_{12}^{-1} \frac{\partial W}{\partial \phi_2} \right) \nonumber \\
%&= \pm \frac{1}{2} \left( 2 \frac{\partial W}{\partial \phi_1} + 3 \frac{\partial W}{\partial \phi_2} \right) \nonumber \\
\frac{d \phi_1}{dx} &=& \pm \frac{1}{2} \left[ - 2 \gamma_1 \sin \phi_1 + \gamma_2 \sin(\phi_1 - \phi_2) - \gamma_3 \sin(2 \phi_1 - \phi_2) \right]\\
%\frac{d \phi_2}{d x} &= \pm \left( \eta_{21}^{-1} \frac{\partial W}{\partial \phi_1} + \eta_{22}^{-1} \frac{\partial W}{\partial \phi_2} \right) \nonumber \\
%&= \pm \frac{1}{2} \left( 3 \frac{\partial W}{\partial \phi_1} + 6 \frac{\partial W}{\partial \phi_2} \right) \nonumber \\
\frac{d \phi_2}{dx} &=& \pm \frac{1}{2} \left[ - 3 \gamma_1 \sin \phi_1 + 3 \gamma_2 \sin(\phi_1 - \phi_2) \right].
\end{eqnarray}

These equations must be solved numerically and some results for the choice\footnote{Different values for these parameters were also considered, however, we observed that the general properties of the solutions did not differ too much from model to model.} $\gamma_i=1$, for $i=1,2,3$ are presented in what follows. 

The numerical solution was based on a discretisation of the system of differential equations (\ref{eq: bps-g2}) and the spatial evolution of the field value is considered using the Runge-Kutta 4 scheme in both directions $x \rightarrow \pm \infty$ from an initial point $\varphi(0) = (\phi_1(0), \phi_2(0))$ which is not one of the points of vacua. The flow of the $\eta$-gradient is then uniquely defined and we should get a non-trivial configuration which connects two vacua of the potential passing through $\varphi(0)$. The mesh discretization is considered to the order of $10^{-5}$ and in order to establish a criterium for the accuracy of the solution we have looked at the difference between the value of the static energy of the numerical solution and the value of the topological charge which can be obtained theoretically. For all the cases presented here these values matched within the precision of $10^{-10}$.

\begin{figure}[h!]
	\centering
	\begin{subfigure}[t]{0.49\textwidth} % width of left subfigure
		\includegraphics[width=\textwidth]{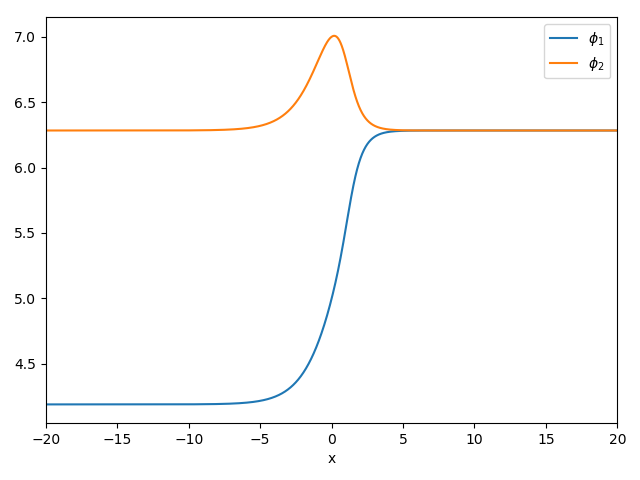}
		\caption{.} % subcaption
		\label{fig: g2 bps 1}
	\end{subfigure}
	\vspace{1em} % here you can insert horizontal or vertical space
	\begin{subfigure}[t]{0.49\textwidth} % width of right subfigure
		\includegraphics[width=\textwidth]{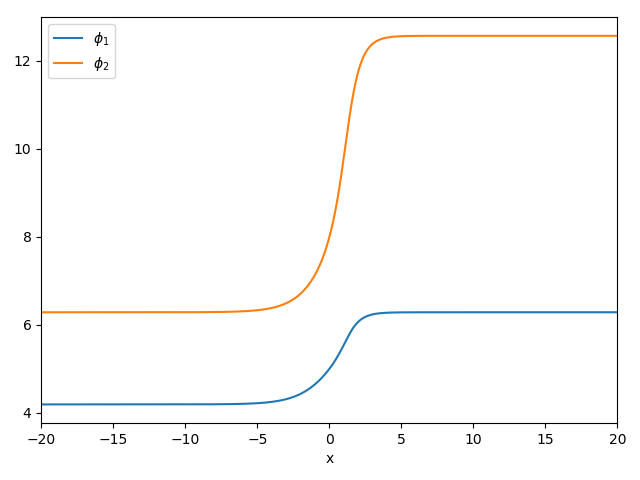}
		\caption{.} % subcaption
		\label{fig: g2 bps 2}
	\end{subfigure}
	\caption{Two numerically calculated BPS states of the model with $\gamma_i = 1$.} % caption for whole figure
\label{fig: g2 bps sol}
\end{figure}

\begin{figure}[h!]
	\centering
		\includegraphics[width=0.49\textwidth]{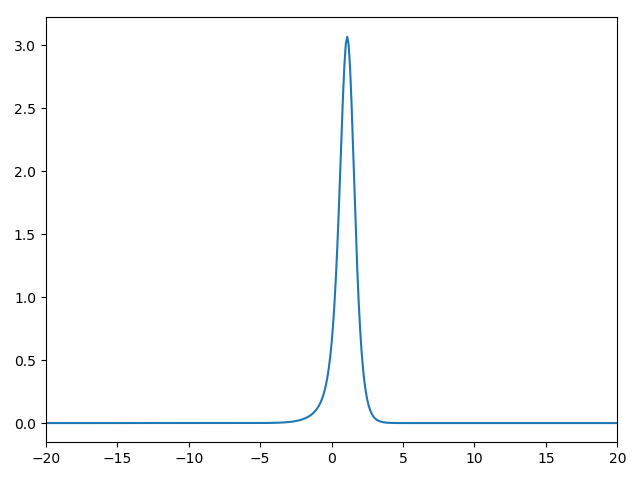}
		\caption{The energy density for both solutions shown in figure \ref{fig: g2 bps 1}. In particular, the energy density for the other solution, \ref{fig: g2 bps 2}, is very much alike.} % subcaption
		\label{fig: g2-engy-dens}
\label{fig: engy-dens-2a}
\end{figure}

In figure \ref{fig: g2 bps sol} we have particular solutions of the BPS equations obtained for different initial conditions $\varphi(0)$ which are close to each other. The solution in figure \ref{fig: g2 bps 1} was constructed from $\varphi(0) = (5, 7)$, \textit{i.e.}, as explained above, it is a solution which interpolates two vacua and passes through the point $\varphi(0)$. The vacua are then chosen by the evolution of the equation. Here we observe that while for the field $\phi_1$ we have the usual kink profile as a BPS state, for $\phi_2$ one finds a single bump. If analysing the two profiles independently, one could have the false impression that $\phi_2$ is topologically trivial. Nevertheless it is important to reinforce the idea that the two fields must not be taken separately, as they are just components of the fundamental field $\varphi$. It is the doublet that contains the topological features of the field and any other relevant physical information. Moreover, note that the configuration as a whole interpolates between vacua $(4 \pi / 3, 2 \pi)$ and $(2 \pi, 2 \pi)$, \emph{i.e.}, the configuration has indeed non-trivial topological data, with topological charge $Q = 9/2$. The energy density, shown in figure \ref{fig: engy-dens-2a} is localized in space, as expected for a solitonic configuration. 

In figure \ref{fig: g2 bps 2} we present another BPS state which was found from the initial value for the field $\phi(0) = (5, 8)$. As a result we have obtained usual kink profiles for both $\phi_1$ and $\phi_2$. This time the solution interpolates between vacua $(4 \pi / 3, 2 \pi)$ and $(2 \pi, 4 \pi)$ and have topological charge $Q = 9/2$ as well.

\begin{figure}[h!]
	\centering
	\begin{subfigure}[t]{0.49\textwidth} % width of left subfigure
		\includegraphics[width=\textwidth]{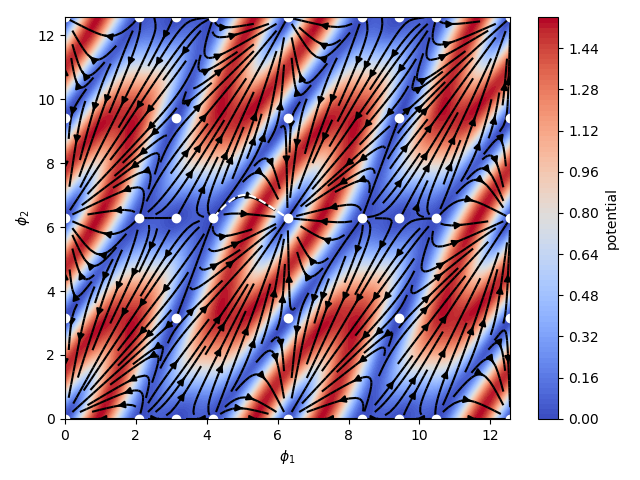}
		\caption{The solution given in (\ref{fig: g2 bps 1}) connects the vacua $(\frac{4\pi}{3},2\pi)$ and $(2\pi,2\pi)$.} % subcaption
		\label{fig: g2 bps 1 pot}
	\end{subfigure}
	\vspace{1em} % here you can insert horizontal or vertical space
	\begin{subfigure}[t]{0.49\textwidth} % width of left subfigure
		\includegraphics[width=\textwidth]{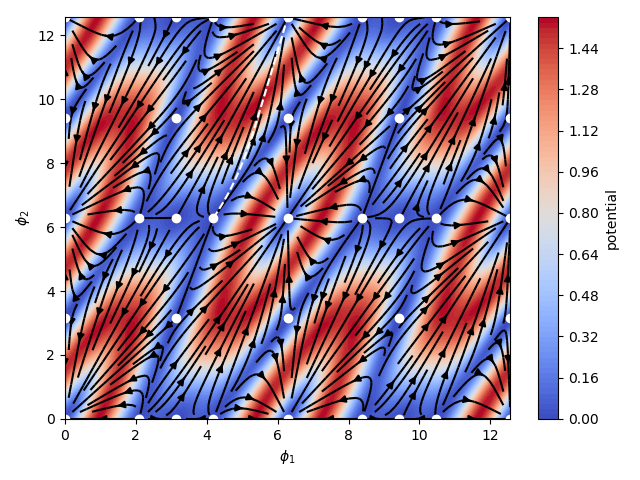}
		\caption{The solution given in (\ref{fig: g2 bps 2}) connects the vacua $(\frac{4\pi}{3},2\pi)$ and $(2\pi,4\pi)$.} % subcaption
		\label{fig: g2 bps 2 pot}
	\end{subfigure}
	\caption{The $\nabla_\eta W$ lines are plotted as black arrows over the potential $U$. The paths described by the solutions in figure \ref{fig: g2 bps sol} are plotted as white dashed lines connecting the vacua which are the white dots.} % caption for whole figure
\label{fig: g2 gradient 1}
\end{figure}

In order to see more clearly the behavior of the doublet $\varphi$ it is quite helpful to look at it as a curve in the field space. This is shown in figures \ref{fig: g2 gradient 1} where we have the $\eta$-gradient lines of the prepotential $W$ plotted, over the potential $U$, in colors. The white dashed lines indicate the BPS solutions given in figure \ref{fig: g2 bps sol}. One can clearly see that the paths follow the $\eta$-gradient flow. Here we can also understand why the configurations in \ref{fig: g2 bps sol} interpolate different vacua. At the point $\phi = (5, 7)$ the $\eta$-gradient flow connects the vacua $(4 \pi / 3, 2 \pi)$ and $(4 \pi / 3, 2 \pi)$, which lie on the same horizontal line $\phi_2 = 2 \pi$. So, from this point of view, it is expected for the profile of $\phi_2$ to have a bump, since it will have to return to the same value at $x = \infty$.

\begin{figure}[h!]
	\centering
	\begin{subfigure}[t]{0.49\textwidth} % width of right subfigure
		\includegraphics[width=\textwidth]{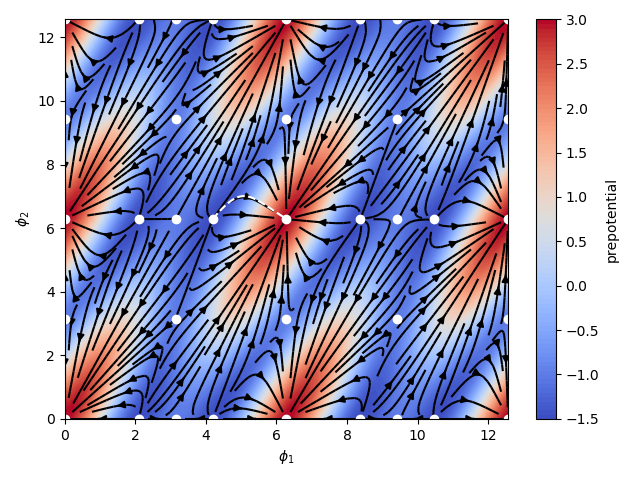}
		\caption{A solution interpolating the vacua $(\frac{4\pi}{3},2\pi)$ and $(2\pi,2\pi)$.} % subcaption
		\label{fig: g2 bps 1 pp}
	\end{subfigure}
	\vspace{1em} % here you can insert horizontal or vertical space
	\begin{subfigure}[t]{0.49\textwidth} % width of right subfigure
		\includegraphics[width=\textwidth]{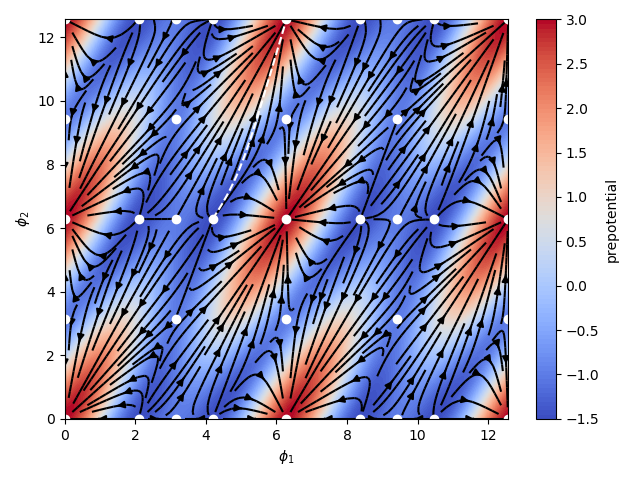}
		\caption{A solution interpolating the vacua $(\frac{4\pi}{3},2\pi)$ and $(2\pi,4\pi)$.} % subcaption
		\label{fig: g2 bps 2 pp}
	\end{subfigure}
	\caption{The $\nabla_\eta W$ lines are plotted as black arrows over the prepotential $W$. The paths described by the solutions in figure \ref{fig: g2 bps sol} are plotted as white dashed lines connecting the vacua which are the white dots.} % caption for whole figure
\label{fig: g2 gradient 2}
\end{figure}

In figure \ref{fig: g2 gradient 2}, the same paths are shown but now on top of the prepotential function $W$, in colors. One can see from this plot that indeed the prepotential evaluated over the BPS solution is a monotonic function of the spatial coordinates $x$ and therefore, the vacuum values are really the points of extrema of this function.

\begin{figure}[h!]
	\centering
	\begin{subfigure}[t]{0.49\textwidth} % width of right subfigure
		\includegraphics[width=\textwidth]{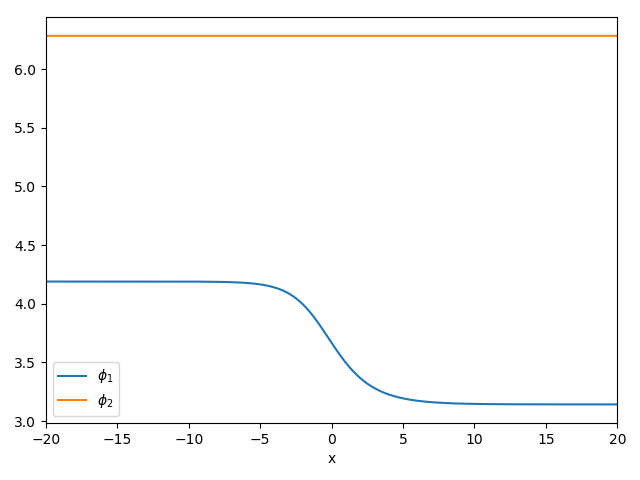}
		\caption{.} % subcaption
		\label{fig: g2 bps 3}
	\end{subfigure}
	\begin{subfigure}[t]{0.49\textwidth} % width of right subfigure
		\includegraphics[width=\textwidth]{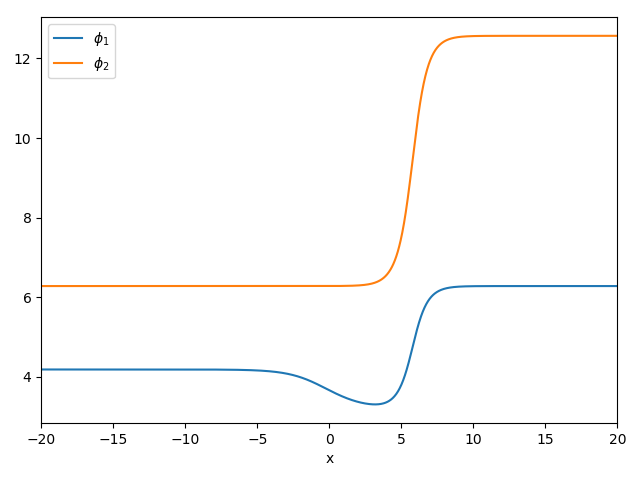}
		\caption{.} % subcaption
		\label{fig: g2 bps 4}
	\end{subfigure}
	\caption{Two numerically calculated BPS states of the model $\gamma_i = 1$.} % caption for whole figure
\label{fig: g2 bps sol2}
\end{figure}

An even more dramatic change in profiles was obtained for the initial condition $\varphi(0) = (7 \pi / 6, 2 \pi)$, in figure \ref{fig: g2 bps 3}, and for $\varphi(0) = (7 \pi / 6, 2 \pi + 10^{-3})$, in figure \ref{fig: g2 bps 4}. Here we see that even a slight change of order $10^{-3}$ in the initial condition $\varphi(0)$ was enough to completely change the solution and the vacua it interpolates. This evidenciates the infinite number of independent BPS solutions there are in this model. In figure \ref{fig: g2 bps 3}, while $\phi_1$ has a kink profile, $\phi_2$ remains constant and the configuration interpolates points $(4 \pi / 3, 2 \pi)$ and $(\pi, 2 \pi)$. This solution has topological charge $Q = 9/2$. In figure \ref{fig: g2 bps 4} we have basically two kink profiles interpolating $(4 \pi / 3, 2 \pi)$ and $(2 \pi, 4 \pi)$, with the exception that $\phi_1$ presents a little bump before tunneling to the other vacuum. The topological charge of this solution is $Q=1/2$.

\begin{figure}[h!]
	\centering
	\begin{subfigure}[t]{0.49\textwidth} % width of left subfigure
		\includegraphics[width=\textwidth]{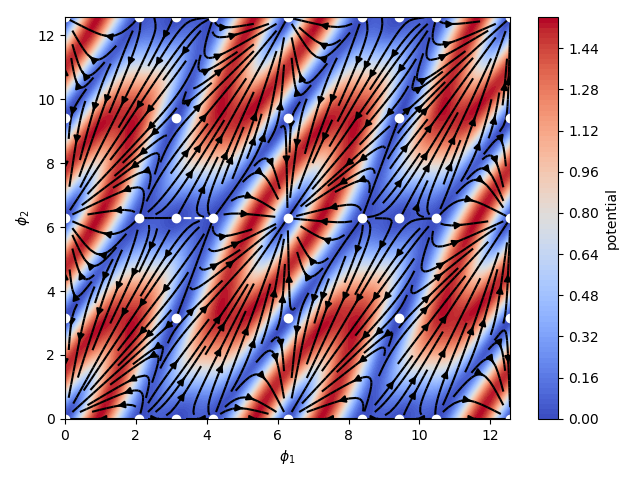}
		\caption{A solution interpolating the vacua $(\frac{4\pi}{3},2\pi)$ and $(\pi,2\pi)$.} % subcaption
		\label{fig: g2 bps pathpot-3}
	\end{subfigure}
	\vspace{1em} % here you can insert horizontal or vertical space
	\begin{subfigure}[t]{0.49\textwidth} % width of right subfigure
		\includegraphics[width=\textwidth]{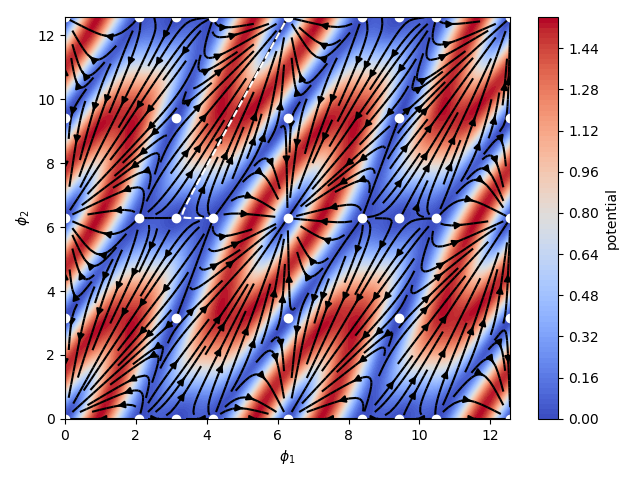}
		\caption{A solution interpolating the vacua $(\frac{4 \pi}{3}, 2 \pi)$ and $(2 \pi, 4 \pi)$.} % subcaption
		\label{fig: g2 bps pathpot-4}
	\end{subfigure}
	\caption{BPS states constructed using slightly different initial conditions. A small change of order $10^{-3}$ in the value of $\phi_2$ implies a dramatic change in the resulting configurations.} % caption for whole figure
\label{fig: g2 gradient 3}
\end{figure}

In figures \ref{fig: g2 gradient 3} we have shown the paths described by the configurations in \ref{fig: g2 bps 3} and \ref{fig: g2 bps 4}, on top of the potential color plot. Notice that the vacua $(\frac{4\pi}{3},2\pi)$ and $(\pi,2\pi)$ are connected exactly by one $\eta$-gradient line. Any variation, no matter how small, in vertical axis, $\phi_2$ would go to a region where the $\eta$-gradient flows opposite to vacuum $(\pi,2\pi)$. This is exactly what was observed in the solutions shown in figures \ref{fig: g2 bps pathpot-3} and \ref{fig: g2 bps pathpot-4}.

\begin{figure}[h!]
	\centering
	\begin{subfigure}[t]{0.49\textwidth} % width of left subfigure
		\includegraphics[width=\textwidth]{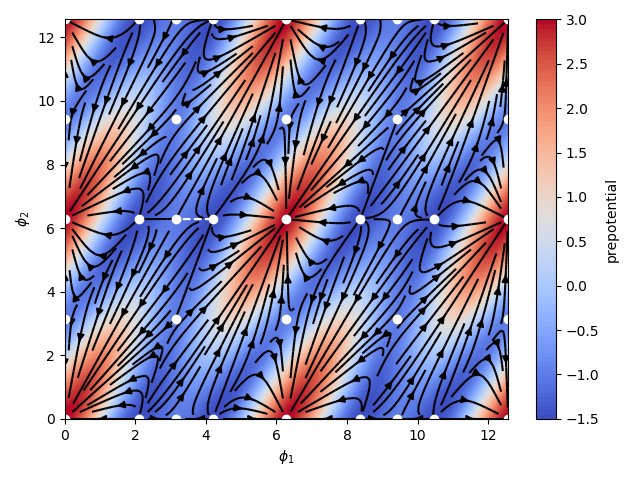}
		\caption{A solution interpolating the vacua $(\frac{4\pi}{3},2\pi)$ and $(\pi,2\pi)$.} % subcaption
		\label{fig: g2 bps pathpp-3}
	\end{subfigure}
	\vspace{1em} % here you can insert horizontal or vertical space
	\begin{subfigure}[t]{0.49\textwidth} % width of right subfigure
		\includegraphics[width=\textwidth]{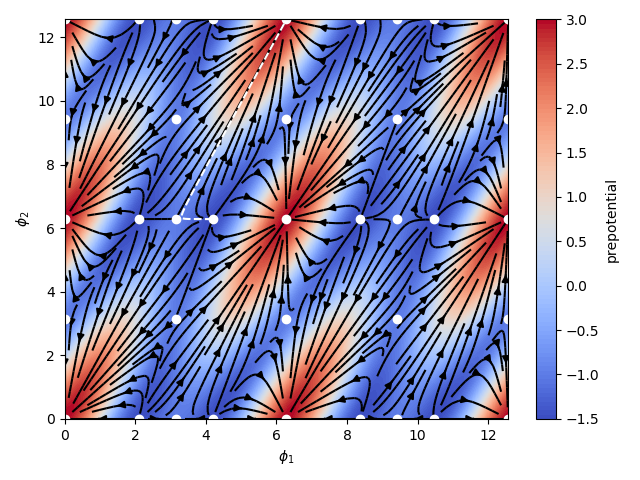}
		\caption{A solution interpolating the vacua $(\frac{4 \pi}{3}, 2 \pi)$ and $(2 \pi, 4 \pi)$.} % subcaption
		\label{fig: g2 bps pathpp-4}
	\end{subfigure}
	\caption{The curves in the field space corresponding to the BPS solutions are shown on top of the prepotential color plot.} % caption for whole figure
\label{fig: g2 gradient 4}
\end{figure}

In figure \ref{fig: g2 gradient 4} we present the same curves obtained from the BPS solutions, but now seen on top of the prepotential colour plot, in order to emphasize that the function $W$ evaluated on the solution is monotonic in $x$.

\subsubsection{The sine-Gordon submodel}

This model has a particular set of solutions which coincide with BPS solutions of the sine-Gordon theory\footnote{Here we discuss the case with $\gamma_1=\gamma_2=\gamma_3=1$. The most general scenario is that with $\gamma_2 = \gamma_3$.}.They are obtained by taking $\phi_1=2\pi n$, with $n\in \mathbb{Z}$. For this choice, the r.h.s of BPS equation for $\phi_1$ vanishes and therefore $\phi_1$ remains constant\footnote{Since the $\phi_1$ component will not play a role in the model anymore, we can reagard this choice as characterizing a submodel of the $\mathfrak{g}_2$ FKZ model.} while the BPS equation for $\phi_2$ is equivalent to that of the sine-Gordon equation\footnote{One can rescale the field and the spatial coordinate in order to get the usual sine-Gordon BPS equation.}
\begin{equation}
\frac{d\phi_2}{dx}=\mp \frac{3}{2}\sin\phi_2
\end{equation}

\begin{figure}[h!]
	\centering
	\begin{subfigure}[t]{0.49\textwidth} % width of left subfigure
		\includegraphics[width=\textwidth]{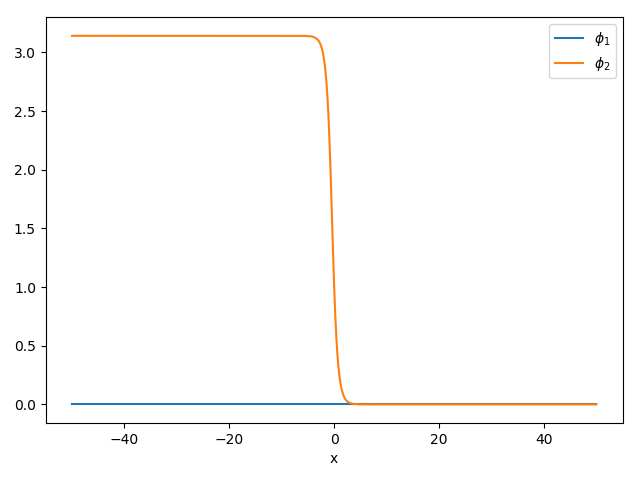}
		\caption{The profile for the case $\varphi=(0,\phi_2)$, where $\phi_2$ satisfies a sine-Gordon equation.} % subcaption
		\label{fig: g2 sg prof}
	\end{subfigure}
	\vspace{1em} % here you can insert horizontal or vertical space
	\begin{subfigure}[t]{0.49\textwidth} % width of right subfigure
		\includegraphics[width=\textwidth]{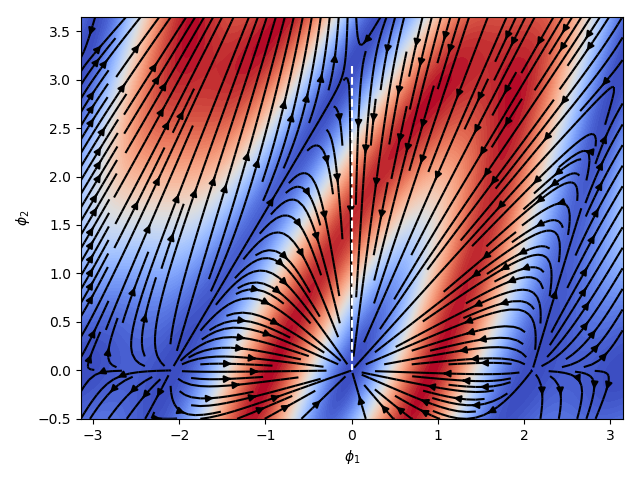}
		\caption{When $\phi_1=0$ and $\phi_2$ is a sine-Gordon antikink, the path in the fields space is a straigh line connecting two vacua.} % subcaption
		\label{fig: g2 sg path}
	\end{subfigure}
	\caption{The sine-Gordon theory appears as a submodel of the $\mathfrak{g}_2$ FKZ model.} % caption for whole figure
\label{fig: g2 sg}
\end{figure}

The profile of the solution in this case is presented in figure \ref{fig: g2 sg} together with the path in the field space. Time dependent solutions of this submodel can be obtained by considering Lorentz boosts of the field: $\varphi(x)\rightarrow \varphi(\frac{x-vt}{\sqrt{1-v^2}})$. This will correspond to a moving profile of $\phi_2$ but in field space the straight line connecting the vacua shown in fugure \ref{fig: g2 sg path} remains unchanged; a boost is equivalent to a reparameterization of this curve.

It is clear that there is an infinite number of paths connecting these two vacua following the $\eta$-gradient flow, all of them described by BPS states with exactly the same energy and topological data. We do not have, in principle, any topological or energetic arguments demonstrating a preference for a specific choice of path, \textit{i.e.} for why the system chooses one solution and not the other amongst the infinitly many of them connecting the same two vacua. For the four solutions presented here, three of them have the same topological charge and exhibit a very different field profile and not only that but even the vacua which are interpolated by solutions of same topological charges can be different as in the case of the solutions presented in figure \ref{fig: g2 bps sol}.

Given the apparent degeneracy in energy one could also think about the possibility of one BPS state evolving in time to another one as a result of some instability. We have performed the numerical time evolution of the dynamical equation, \textit{i.e.}, the evolution of the BPS configuration using the full dynamical equation (\ref{eq:eom}), and our results have shown that these BPS configurations remain very stable. This seems to be corroborated by the fact that being a BPS state a state of lowest possible energy, its change to a different state would require a spare energy amount which does not exist for the transition to happen. The fact that the numerically calculated energy agrees with a very good precision to the topological charge indicates that we are certain that the numerical value of the field is close enough to its constant value expected at infinity and no remaining non-zero spatial derivative term exists there, which could lead to a gain in the energy as the result of a numerical artifact.

%%%%%%%%%%%%%%%%%%%%%%%%%%%%%%%%%%%%%%%%%%%%%%%%%%%%%%%%%%%%%%%%%%%%%%%%%%%5

\section{The FKZ model for the algebra $\mathfrak{su}(4)$}
\label{sec:su4}
\setcounter{equation}{0}

\subsection{The construction of the model}

The algebra $\mathfrak{su}(4)$ is of rank $r = 3$ and the FKZ model associated to it describes a model for a scalar triplet:
\begin{align}
\varphi = \phi_1 \frac{2 \alpha_1}{\norm{\alpha_1}^2} + \phi_2 \frac{2 \alpha_2}{\norm{\alpha_2}^2} + \phi_3 \frac{2 \alpha_3}{\norm{\alpha_3}^2}.
\end{align}
The Cartan matrix of $\mathfrak{su}(4)$ is
\begin{align}
K = \left( \begin{array}{ccc}
2 & -1 & 0 \\
-1 & 2 & -1 \\
0 & -1 & 2
\end{array} \right),
\end{align}
and taking $\norm{\alpha_a}^2=1$ the field space metric is taken to be
\begin{align}
\eta = \left( \begin{array}{ccc}
4 & -2 & 0 \\
-2 & 4 & -2 \\
0 & -2 & 4
\end{array} \right).
\end{align}

Again, the starting point for the construction of the FKZ model is to define the prepotential, that is, to define the representation of the algebra. The algebra $\mathfrak{su}(4)$ has three fundamental weights given by
\begin{align}
\lambda_1 &= \frac{1}{4} \left( 3 \alpha_1 + 2 \alpha_2 + \alpha_3 \right) \nonumber \\
\lambda_2 &= \frac{1}{2} \left( \alpha_1 + 2 \alpha_2 + \alpha_3 \right) \\
\lambda_3 &= \frac{1}{4} \left( \alpha_1 + 2 \alpha_2 + 3\alpha_3 \right).\nonumber 
\end{align}

For the first fundamental representation the weights are
\begin{align}
\mu_1 &= \lambda_1 = \frac{1}{4} \left( 3 \alpha_1 + 2 \alpha_2 + \alpha_3 \right) \nonumber \\
\mu_2 &= \lambda_1 - \alpha_1 = \frac{1}{4} \left( - \alpha_1 + 2 \alpha_2 + \alpha_3 \right) \nonumber \\
\mu_3 &= \lambda_1 - \alpha_1 - \alpha_2 = \frac{1}{4} \left( - \alpha_1 - 2 \alpha_2 + \alpha_3 \right) \\
\mu_4 &= \lambda_1 - \alpha_1 - \alpha_2 - \alpha_3 = - \frac{1}{4} \left( \alpha_1 + 2 \alpha_2 + 3 \alpha_3 \right), \nonumber 
\end{align}
and those of the second fundamental representation are
\begin{align}
\tilde{\mu}_1 &= \lambda_2 = \frac{1}{2} \left( \alpha_1 + 2 \alpha_2 + \alpha_3 \right) \nonumber \\
\tilde{\mu}_2 &= \lambda_2 - \alpha_2 = \frac{1}{2} \left( \alpha_1 + \alpha_3 \right) \nonumber \\
\tilde{\mu}_3 &= \lambda_2 - \alpha_1 - \alpha_2 = \frac{1}{2} \left( - \alpha_1 + \alpha_3 \right) \\
\tilde{\mu}_4 &= \lambda_2 - \alpha_2 - \alpha_3 = \frac{1}{2} \left( \alpha_1 - \alpha_3 \right) = - \tilde{\mu}_3 \nonumber \\
\tilde{\mu}_5 &= \lambda_2 - \alpha_1 - \alpha_2 - \alpha_3 = - \frac{1}{2} \left( \alpha_1 + \alpha_3 \right) = - \tilde{\mu}_2 \nonumber 
\end{align}
and, finaly, the weights of the third fundamental representation are
\begin{align}
\bar{\mu}_1 &= \lambda_3 = \frac{1}{4} \left( \alpha_1 + 2 \alpha_2 + 3 \alpha_3 \right) = - \mu_4 \nonumber \\
\bar{\mu}_2 &= \lambda_3 - \alpha_3 = \frac{1}{4} \left( \alpha_1 + 2 \alpha_2 - \alpha_3 \right) = - \mu_3 \nonumber \\
\bar{\mu}_3 &= \lambda_3 - \alpha_3 - \alpha_2 = \frac{1}{4} \left( \alpha_1 - 2 \alpha_2 - \alpha_3 \right) = - \mu_2 \\
\bar{\mu}_4 &= \lambda_3 - \alpha_3 - \alpha_2 - \alpha_1 = - \frac{1}{4} \left( 3 \alpha_1 + 2 \alpha_2 + \alpha_3 \right) = - \mu_1. \nonumber 
\end{align}

None of the fundamental representations alone satisfy the requirement for reality of the prepotential, but we notice that a direct sum of the first and third, $4 \oplus \bar{4}$, does. As we only need one of each pair of weights in order to construct the prepotential, we use the weights of representation $4$ and calculate the internal products with the field $\varphi$:
\begin{align}
\mu_1 \cdot \varphi &= \sum_{a = 1}^3 \phi_a \frac{2 \lambda_1 \cdot \alpha_a}{\norm{\alpha_a}^2} = \sum_{a = 1}^3 \phi_a \delta_{1a} = \phi_1 \nonumber \\
\mu_2 \cdot \varphi &= \sum_{a = 1}^3 \phi_a \frac{2 (\lambda_1 - \alpha_1) \cdot \alpha_a}{\norm{\alpha_a}^2} = \phi_1 - K_{11} \phi_1 - K_{12} \phi_2 - K_{13} \phi_3 = \phi_2 - \phi_1 \nonumber \\
\mu_3 \cdot \varphi &= \sum_{a = 1}^3 \phi_a \frac{2 (\lambda_1 - \alpha_1 - \alpha_2) \cdot \alpha_a}{\norm{\alpha_a}^2} = \phi_2 - \phi_1 - K_{21} \phi_1 - K_{22} \phi_2 - K_{23} \phi_3 = \phi_3 - \phi_2 \\
\mu_4 \cdot \varphi &= \sum_{a = 1}^3 \phi_a \frac{2 (\lambda_1 - \alpha_1 - \alpha_2 - \alpha_3) \cdot \alpha_a}{\norm{\alpha_a}^2} = \phi_3 - \phi_2 - K_{31} \phi_1 - K_{32} \phi_2 - K_{33} \phi_3 = - \phi_3. \nonumber 
\end{align}

Then, the prepotential for the $\mathfrak{su}(4)$ FKZ model for the representation $4 \oplus \bar{4}$ is given by
\begin{align} \label{eq: su4 prepot}
W = \gamma_1 \cos \phi_1 + \gamma_2 \cos(\phi_1 - \phi_2) + \gamma_3 \cos(\phi_2 - \phi_3) + \gamma_4 \cos \phi_3,
\end{align}
and the components of the gradient of the prepotential in the field space are
\begin{align} \label{eq: su4 grad}
\frac{\partial W}{\partial \phi_1} &= - \gamma_1 \sin \phi_1 - \gamma_2 \sin(\phi_1 - \phi_2) \nonumber \\
\frac{\partial W}{\partial \phi_2} &= \gamma_2 \sin(\phi_1 - \phi_2) - \gamma_3 \sin(\phi_2 - \phi_3) \\
\frac{\partial W}{\partial \phi_3} &= \gamma_3 \sin(\phi_2 - \phi_3) - \gamma_4 \sin \phi_3. \nonumber 
\end{align}

The potential for this model is then written in terms of these components as
\begin{align} \label{eq: su4 pot}
U(\varphi) &= \frac{1}{2} \left[ \frac{3}{8} \left( \frac{\partial W}{\partial \phi_1} \right)^2 + \frac{1}{2} \left( \frac{\partial W}{\partial \phi_2} \right)^2 + \frac{3}{8} \left( \frac{\partial W}{\partial \phi_3} \right)^2 + \frac{1}{2} \frac{\partial W}{\partial \phi_1} \frac{\partial W}{\partial \phi_2} + \frac{1}{4} \frac{\partial W}{\partial \phi_1} \frac{\partial W}{\partial \phi_3} + \frac{1}{2}\frac{\partial W}{\partial \phi_2} \frac{\partial W}{\partial \phi_3} \right],
\end{align}
which, for the choice $\gamma_i=1$, $i=1,2,3$, adopted from now on, reads
\begin{align}
U(\varphi)&=\frac{3}{16} \sin ^2\left(\phi _1\right)+\frac{3}{16} \sin ^2\left(\phi _1-\phi _2\right)+\frac{3}{16} \sin ^2\left(\phi _2-\phi _3\right)+\frac{3}{16} \sin ^2\left(\phi _3\right)+\nonumber\\
&+\left(\frac{1}{8} \sin \left(\phi _1-\phi _2\right)+\frac{1}{8} \sin \left(\phi _2-\phi _3\right)+\frac{\sin \left(\phi _3\right)}{8}\right) \sin \left(\phi _1\right)+\nonumber\\
&-\left(\frac{1}{8} \sin \left(\phi _2-\phi _3\right)\frac{\sin \left(\phi _3\right)}{8}\right) \sin \left(\phi _1-\phi _2\right)-\frac{1}{8} \sin \left(\phi _2-\phi _3\right) \sin \left(\phi _3\right).
\end{align}

\begin{figure}[h!]
	\centering
		\includegraphics[width=0.6\textwidth]{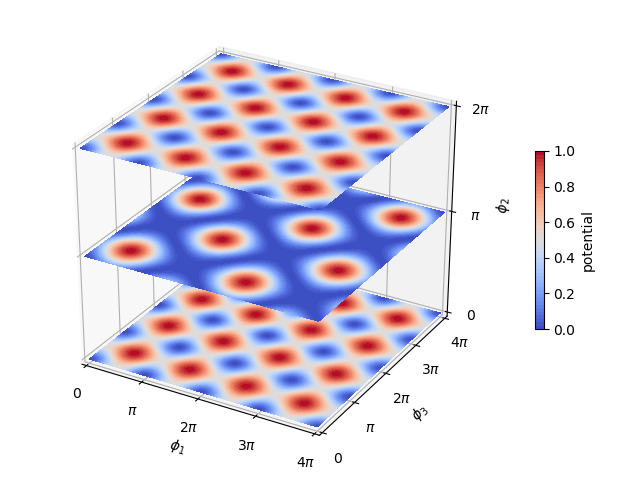}
		\caption{The potential energy density is seen in sliced planes $\phi_2 = 0$, $\phi_2=\pi$ and $\phi_2=2\pi$. } % subcaption
	\label{fig: pot_slice}
\end{figure}

This potential function can be visualized in figure \ref{fig: pot_slice}.

\subsection{BPS solutions}

\begin{figure}[h!]
	\centering
		\includegraphics[width=0.6\textwidth]{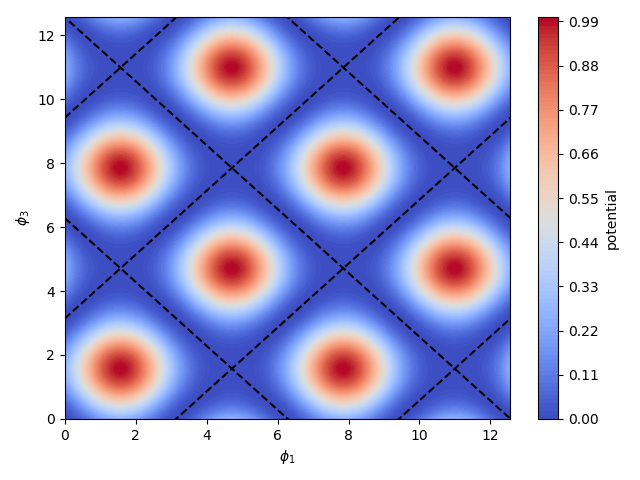}
		\caption{The potential $U(\phi_1,\phi_3) = \frac{1}{4}(\sin{\phi_1}+\sin{\phi_3})^2$, restricted to a plane $\phi_2 = (2n+1)\pi$, $n\in \mathbf{Z}$. The vacua lines $\phi_3 =  \phi_1 + (2n+1)\pi$, and $\phi_3 = -\phi_1+2n\pi$, $n \in \mathbb{Z}$ are shown. } % subcaption
	\label{fig: potplane}
\end{figure}

The vacuum manifold is defined by the set of points of minima of the potential. These points are the solutions of the system of equations (\ref{eq:vaccond}). For our choice $\gamma_i = 1$, $i=1,2,3$, besides the usual discrete set of vacua defined by integer multiples of $\pi$, this system has also solutions given by straight lines in the field space, $\phi_3 =  \phi_1 + (2n+1)\pi$, and $\phi_3 = -\phi_1+2n\pi$, $n \in \mathbb{Z}$, defined on the planes $\phi_1\phi_3$ which are located at $\phi_2 = (2m+1)$, $m\in \mathbb{Z}$. This continuous set of points, referred here as ``vacua lines'' can be seen in figure \ref{fig: potplane}.

This model has also some discrete symmetries, namely, $(\phi_1,\phi_2, \phi_3) \rightarrow (\phi_3,\phi_2, \phi_1)$, $\phi_a \rightarrow \phi_a + 2n_a \pi$ and $(\phi_1,\phi_2, \phi_3) \rightarrow (\phi_1 + n\pi,\phi_2, \phi_3+ n\pi)$. This means that the set of vacuum points defined above can be further extended including these transformations. Using a parameter $\tau$ in the field space, and taking into account these symmetries, the vacuum manifold can be finally defined as
\begin{equation}
\varphi = (n_1\pi, n_2\pi, n_3\pi), \qquad \qquad \varphi(\tau) = (\tau, \pi, \pi + \tau), \qquad \qquad \varphi(\tau) = (\tau, \pi, - \tau),
\end{equation}
with $n_1$, $n_2$ and $n_3$ integers.

\subsubsection{Perturbations on the continuous vacua line}

On the plane $\phi_2 = \pi$ the potential is highly anisotropic except along the directions of the vacua lines, where it remains constant with minimum value. The potential has a translation invariance along these particular directions. Given a point $\varphi_0 = (\tau, \pi, \tau + \pi)$, where $\tau$ is fixed, in one of these vacua lines, one can obtain another point of vacuum by performing a continuous translation 
%with a parameter $\epsilon$ $\varphi'=\varphi_0+ \epsilon(1,0,1)$,
along this line. Now, if a perturbation along the parallel and perpendicular directions to the vacua line is considered\footnote{The introduction of a perturbation in the perperndicular direction to the plane where the vacua line is will not contribute to this linear approximation.} as 
\begin{equation}
\varphi = \varphi_0 + \frac{1}{4}(\theta - \chi, 0 , \theta + \chi),
\end{equation}
with $\theta$ and $\chi$ the perturbation fields along these respective directions, then the lagrangian up to quadratic order in these perturbations reads
\begin{equation}
\mathcal{L}_{\theta,\chi} = \frac{1}{2}\partial_\mu \theta \partial^\mu \theta + \frac{1}{2} \partial_\mu \chi \partial^\mu \chi - \frac{1}{16}\left(\cos^2{\tau}\right)\chi^2
\end{equation}
and the field $\theta$ will be a massless while $\chi$ can be either massless or massive: the mass term is given by $m_{\chi} = \frac{1}{2\sqrt{2}}\cos\tau$, which depends on the choice of the initial vacuum, \textit{i.e.}, on the choice of $\tau$, so that for $\tau = (n+\frac{1}{2})\pi$, $n\in \mathbf{Z}$, which are exactly the points where the two types of vacua lines intersect, $m_\chi =0$.

Although a general translation in the field space is not a symmetry of the model, the potential is invariant for such translations restricted to the continuous vacua line. Nevertheless, the anisotropy of the potential in the perpendicular direction to this line on the plane becomes manifest in the fact that the perturbations in this direction aquire different masses depending on the point along this line they are considered.

After this analysis of the vacua of the potential we can finally solve the BPS equations for our model which are given below:
\begin{eqnarray}
\frac{d \phi_1}{d x} &=& \mp \frac{1}{8} \left(3  \sin \left(\phi _1\right)+ \sin \left(\phi _1-\phi _2\right)+ \sin \left(\phi _2-\phi _3\right)+ \sin \left(\phi _3\right)\right)\\
\frac{d\phi_2}{dx}&=& \pm \frac{1}{4} \left(- \sin \left(\phi _1\right)+ \sin \left(\phi _1-\phi _2\right)- \sin \left(\phi _2-\phi _3\right)- \sin \left(\phi _3\right)\right)\\
\frac{d\phi_3}{dx}&=&\pm \frac{1}{8} \left(- \sin \left(\phi _1\right)+ \sin \left(\phi _1-\phi _2\right)+ \sin \left(\phi _2-\phi _3\right)-3  \sin \left(\phi _3\right)\right).
\end{eqnarray}

\begin{figure}[h!]
	\centering
	\begin{subfigure}[t]{0.49\textwidth} % width of left subfigure
		\includegraphics[width=\textwidth]{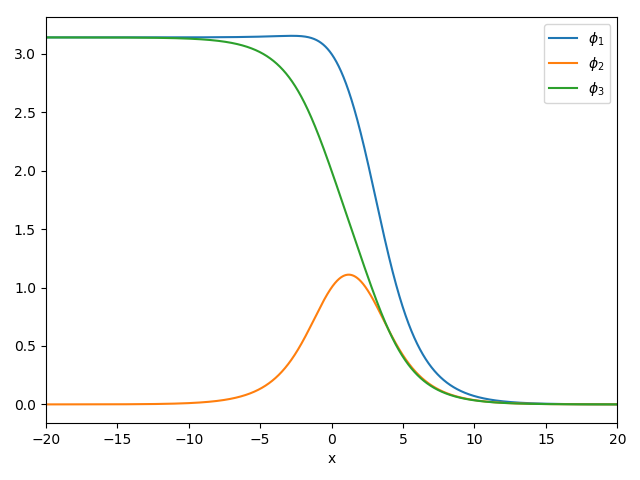}
		\caption{Two of the components of the BPS configuration exhibit a kink-like behaviour while the third one has a different shape.} % subcaption
		\label{fig: su4 bps 1}
	\end{subfigure}
	\vspace{1em} % here you can insert horizontal or vertical space
	\begin{subfigure}[t]{0.49\textwidth} % width of right subfigure
		\includegraphics[width=\textwidth]{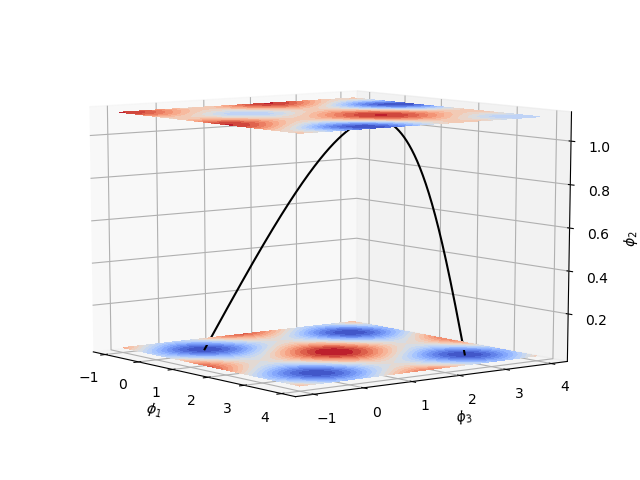}
		\caption{The path in the fields space which defines a BPS state connects two vacua of the potential} % subcaption
		\label{fig: su4 path 1}
	\end{subfigure}
	\caption{A BPS solution of the $\mathfrak{su}(4)$ FKZ model.}
	\label{fig: su4 1}
\end{figure}

The numerical integration of these equations follows the same scheme as used in the case of two fields, for the $\mathfrak{g}_2$ FKZ model. In figure \ref{fig: su4 1} we present a solution obtained from the starting point $\varphi(0)=(3,1,2)$. The profile of each component of the triplet is shown in figure \ref{fig: su4 bps 1} and the path in the field space is plotted in figure \ref{fig: su4 path 1}, together with slices of the potential in planes of constant $\phi_2$. One sees that the path connects two minima of the potential, as expected.

\subsubsection{The sine-Gordon submodels}

\begin{figure}[h!]
	\centering
	\begin{subfigure}[t]{0.49\textwidth} % width of left subfigure
		\includegraphics[width=\textwidth]{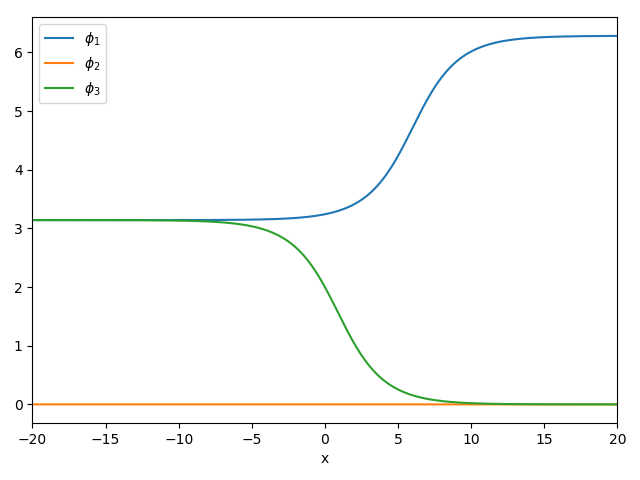}
		\caption{The profile of the components of the scalar triplet obtained with $\varphi(0)=(\pi+0.1,0,2)$.} % subcaption
		\label{fig: su4 bps 3}
	\end{subfigure}
	\vspace{1em} % here you can insert horizontal or vertical space
	\begin{subfigure}[t]{0.49\textwidth} % width of right subfigure
		\includegraphics[width=\textwidth]{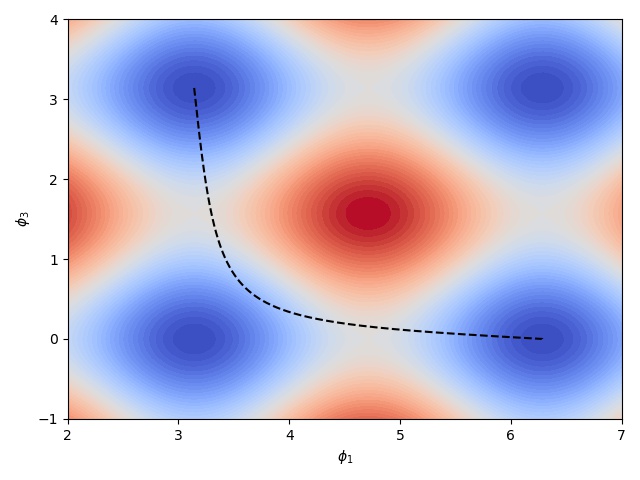}
		\caption{The path in the fields space remains in the plane $\phi_2=0$.} % subcaption
		\label{fig: su4 path 3}
	\end{subfigure}
	\caption{For $\phi_2=$ an integer multiple of $2\pi$ the BPS equations become the equations of two independent sine-Gordon models for the components $\phi_1$ and $\phi_3$ of the scalar triplet.}
	\label{fig: su4 2}
\end{figure}

\begin{figure}[h!]
	\centering
	\begin{subfigure}[t]{0.49\textwidth} % width of left subfigure
		\includegraphics[width=\textwidth]{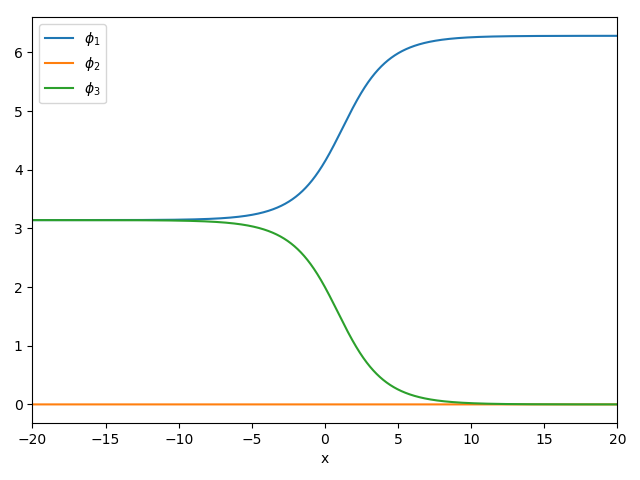}
		\caption{The component $\phi_1$ is shifted in space if compared with the one shown in \ref{fig: su4 bps 3}. } % subcaption
		\label{fig: su4 bps 4}
	\end{subfigure}
	\vspace{1em} % here you can insert horizontal or vertical space
	\begin{subfigure}[t]{0.49\textwidth} % width of right subfigure
		\includegraphics[width=\textwidth]{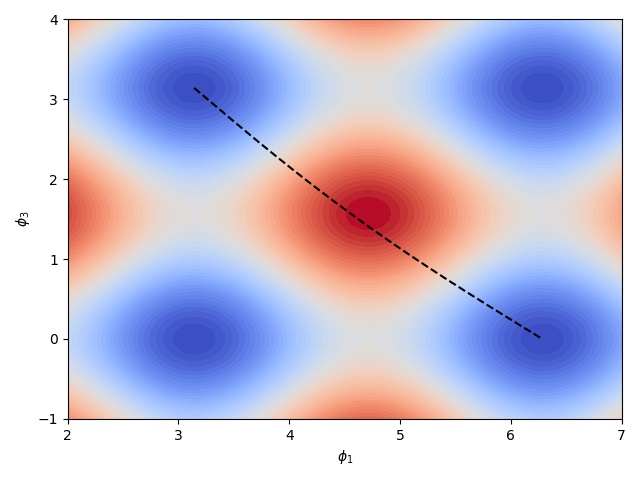}
		\caption{The path in the fields space has completely changed from the one obtained in \ref{fig: su4 path 3}.} % subcaption
		\label{fig: su4 path 4}
	\end{subfigure}
	\caption{A BPS solution of the $\mathfrak{su}(4)$ FKZ model for the particular case where $\phi_2=0$.}
	\label{fig: su4 3}
\end{figure}

Next we have some solutions for which $\phi_2 = 0$. In this case the r.h.s. of the BPS equation for $\phi_2$ vanishes and we have two independent sine-Gordon models for $\phi_1$ and $\phi_3$. The profiles of $\phi_1$ and $\phi_3$ are shown in two different cases: one obtained by considering $\varphi(0) = (\pi+0.1,0,2)$, shown in figure \ref{fig: su4 2}, and the other considering $\varphi(0)=(\pi+1,0,2)$, shown in figure \ref{fig: su4 3}. 

From the BPS equations one sees that the $\phi_2$ has a role of coupling the two fields $\phi_1$ and $\phi_3$. By setting it to an integer multiple of $2\pi$, the equations for $\phi_1$ and $\phi_3$ become uncoupled and one has the freedom of shifting these two fields in space independently. So one can scan a whole subset of solutions of the $\mathfrak{g}_2$ FKZ model starting from such a configuration and performing translations of the fields $\phi_1$ and $\phi_3$. If one considers a Lorentz boost which is performed equally for all the components of the scalar triplet, then one finds a class of time dependent solutions.
 
\begin{figure}[h!]
	\centering
	\begin{subfigure}[t]{0.49\textwidth} % width of left subfigure
		\includegraphics[width=\textwidth]{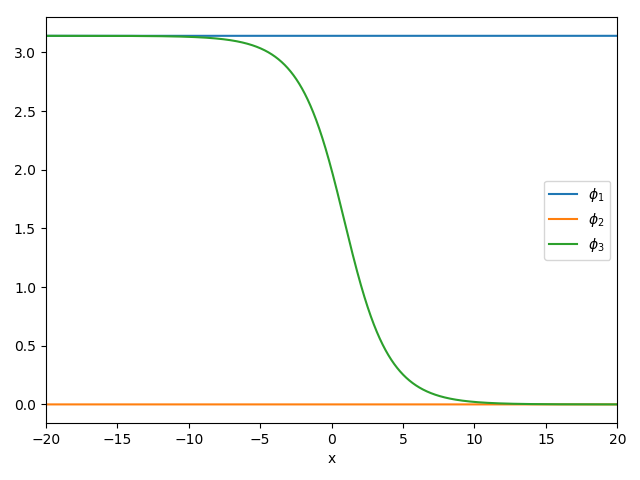}
		\caption{The component $\phi_1$ is shifted in space if compared with the one shown in \ref{fig: su4 bps 3}. } % subcaption
		\label{fig: su4 bps 5}
	\end{subfigure}
	\vspace{1em} % here you can insert horizontal or vertical space
	\begin{subfigure}[t]{0.49\textwidth} % width of right subfigure
		\includegraphics[width=\textwidth]{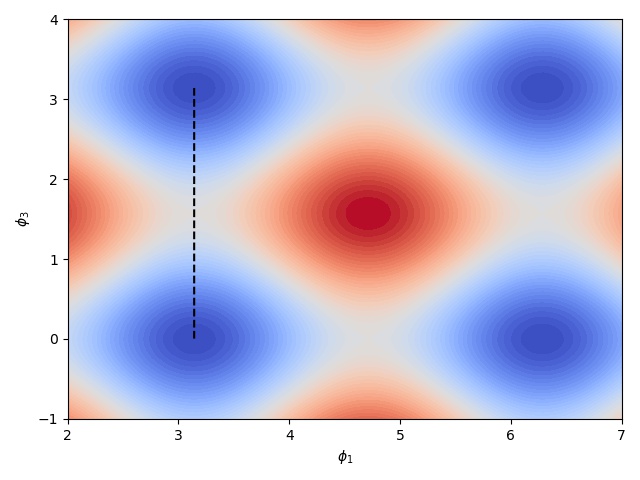}
		\caption{The path in the fields space has completely changed from the one obtained in \ref{fig: su4 path 3}.} % subcaption
		\label{fig: su4 path 5}
	\end{subfigure}
	\caption{A BPS solution of the $\mathfrak{su}(4)$ FKZ model for the particular case where $\phi_2=0$.}
	\label{fig: su4 4}
\end{figure}

Another submodel can be further obtained if one consider $\phi_1$ an integer multiple of $\pi$ together with $\phi_2$ an integer multiple of $2\pi$. In figure \ref{fig: su4 4} we show the case obtained from $\varphi(0)=(\pi,0,2)$. It is a sine-Gordon model for the field $\phi_3$. Here, again, we have a whole class of solutions which can be obtained by shifting this configuration.

We have performed the numerical integration of the full dynamical equations (\ref{eq:eom}) taking the BPS configurations as the starting profile. As expected, the solutions are very stable. 
\section{Conclusions}
\label{sec:conclusions}
\setcounter{equation}{0}

We have explored two new models based on the generalization of the BPS equation introduced in \cite{Ferreira:2018ntx}, where the field space is that of the roots of a given Lie algebra, here considered to be $\mathfrak{g}_2$ and $\mathfrak{su}(4)$. Usually we have solitonic solutions in 1+1 dimensional space-time for theories involving a single scalar field. The models we have discussed here are in 1+1 dimensions but the field is a doublet and a triplet. What we see is that the number of vacua is huge and moreover, the possibilities of solutions which interpolate them is also very large. This allows for the construction of many different configurations which, in the field space, defines a curve or a string with fixed end points. These BPS solutions were proved to be very stable as one could expect for a configuration with the least possible energy. We were not able to provide an argument which justifies the choice of the BPS state made by the system, dynamically, i.e., amongst the infinitely many solutions with same topological charge which interpolate two vacua, our numerical technique leaves for the system to choose one and what we observed is that any of them is as good as the others. Finally, the form of the potential allows for the identification of submodels in the sense that one can make one (or two, in the case of $\mathfrak{su}(4)$) of the components of the multiplet to be a constant and the remaining degrees of freedom will undergo a dynamics governed by the sine-Gordon equation, which allows for the construction of analytical solutions.

We are currently investigating on physical models which can be described by multiplets of scalar field in 1+1 dimensional space-time and therefore, a concrete application of some of the FKZ models, perhaps for specific choices of the parameters $\gamma_i$. We are also solving the full dynamical equation of the solutions found here in order to be able to discuss about the forces between the solitons.

\newpage

\vspace{2cm}

\noindent {\bf Acknowledgments:} The authors would like to thank Luiz Agostinho Ferreira for calling their attention to the paper he wrote in collaboration with Pawel Klimas and Wojtek Zackrewski where the FKZ models were introduced recently. The authors would also like to thank Betti Hartmann for valuable comments on the manuscript.  This  study  was  financed  in  part  by  the  Coordena\c c\~ao  de Aperfei\c coamento  de  Pessoal  de  N\'ivel  Superior  -  Brasil  (CAPES). G. Luchini is supported by FAPES under the EDITAL CNPq/FAPES N$^o$ 22/2018. 
\newpage

%\bibliographystyle{hapalike2-NOand}

%\bibliography{references}

\end{document}